\documentclass[letterpaper,titlepage,11pt]{article}
\usepackage{hyperref}
\usepackage{amssymb,amsmath,amsfonts}
\usepackage{epsfig}
\usepackage{graphicx}
\usepackage{float}
\usepackage[english]{babel}
\usepackage[T1]{fontenc}
\usepackage[applemac]{inputenc}

\def\clock{{\count0=\time
           \divide\count0 60
           \ifnum\count0<10 0\fi\the\count0
           \multiply\count0 -60 \advance\count0 \time
           :\ifnum\count0<10 0\fi \the\count0
         }}
\newcommand{\timestamp}{{\small\vbox{\hbox{\tt\jobname.tex}
\hbox{\the\day/\the\month/\the\year, \clock}}}}

\newtheorem{definition}{Definition}[section]

\newtheorem{conjecture}[definition]{Conjecture}

\newcommand{\spa} {\quad , \quad}
\numberwithin{equation}{section}
\begin{document}

\begin{titlepage}

$\;\;$
\vskip 1.8cm

\centerline{\Huge \bf Blackfolds in (Anti)-de Sitter Backgrounds}

\vskip 1.5cm
\centerline{\bf  Jay Armas and Niels A. Obers}
%\vskip 0.5cm

\vskip 0.8cm
\centerline{\sl The Niels Bohr Institute}
\centerline{\sl Blegdamsvej 17, DK-2100 Copenhagen \O, Denmark}
\vskip 0.6cm

\centerline{\small\tt jay@nbi.dk, obers@nbi.dk}

\vskip 1.2cm

\centerline{\bf Abstract} \vskip 0.2cm \noindent We construct different neutral blackfold solutions in Anti-de Sitter and de Sitter background  spacetimes in the limit where the cosmological constant is taken to be much smaller than the horizon size.
This includes a class of blackfolds with horizons that are products of odd-spheres times a transverse sphere, for which the thermodynamic
stability is also studied. Moreover, we exhibit a specific case in which the same blackfold solution can describe different limiting
black hole spacetimes therefore illustrating the geometric character of the blackfold approach.  Furthermore, we show that the higher-dimensional Kerr-(Anti)-de Sitter black hole allows for ultra-spinning regimes in the same limit under consideration and
 demonstrate that this is correctly described by a pancaked blackfold geometry. We also give evidence for the possibility of saturating the rigidity theorem in these backgrounds.

\vskip 5.5cm

\begin{flushright}
\emph{"He who defends objectivity in science suppresses himself as a subject and is proud of it - considers that such is required for the scientific method. However there are people who do not believe that science can be made by objects."} \\
Gonçalo M. Tavares in \emph{Brief Notes on Science}
\end{flushright}

%\vskip 0.5cm
%\leftline{\timestamp}

\end{titlepage}

\small
%\begin{spacing}{1}
\tableofcontents
%\end{spacing}
%\tableofcontents
\normalsize
%\newpage
%\pagestyle{plain}
\setcounter{page}{1}

%%%%%%%%%%%%%%%%%%%%%%%%%%%%%%%%%%%%%%%%%%%%%%%%%%%%%%%%%%%%%%
%\newpage

\section{Opening remarks}

Recent research  has revealed that the phase structure of higher-dimensional black hole spacetimes is far more complex  and vast\footnote{See Ref.~\cite{Emparan:2008eg} for an extensive review
and \cite{Obers:2008pj}
for shorter reviews of
higher-dimensional black holes.}
than what was \emph{a priori} thought. Since only in special
cases with a high degree of symmetry exact analytic black hole solutions can be obtained,
one is forced to develop new approaches and methodologies in order to probe the intricate space of possible solutions. In particular, if one restricts to asymptotically flat black holes with $D-2$ commuting Killing vectors in four or five dimensions and to gravity theories which have an integrable sector, it is possible to exploit the existent symmetries and develop solution generating techniques. On the other hand, if the spacetime has fewer symmetries, $D\ge6$ or different asymptotics, exact analytic methods are harder to establish.

An approximate analytic method for constructing black hole solutions in any spacetime dimension $D \geq 5$ and for a wide class of backgrounds was recently developed in \cite{Emparan:2009cs,Emparan:2009at}.
The method, called the {\it blackfold approach}, is based on a world-volume effective theory  that describes how to bend the worldvolume of a black brane in a given background spacetime. A systematic scan  of the landscape of black holes in $D\geq 5$  was initiated in \cite{Emparan:2009vd} where large classes of
neutral stationary black holes in a Minkowski background were uncovered, exhibiting novel horizon topologies. These include helical black strings and black rings, black
odd-spheres, for which the horizon is a product of a large and a small
sphere, and non-uniform black cylinders. It was also shown how the blackfold description correctly recovers the
ultra-spinning Myers-Perry (MP) black holes as ellipsoidal even-balls. Earlier  works in which the blackfold methodology
was applied include
\cite{Emparan:2007wm,Caldarelli:2008pz,BlancoPillado:2007iz}
while recently the method
was used to find the thermal generalization of the BIon solution \cite{Grignani:2010xm} by means of the
extension of the blackfold method to charged black branes \cite{upcoming}.

The purpose of this paper is then to begin a systematic scan of similar blackfolds as those found
in  \cite{Emparan:2009vd} but considering
instead embeddings in an (A)dS background. The case of thin black rings in (A)dS space was earlier considered in Ref.~\cite{Caldarelli:2008pz} following the construction of thin black rings in flat space \cite{Emparan:2007wm}.
The former will be shown to be part of a larger family of odd-sphere (A)dS blackfolds, just as the latter were shown in \cite{Emparan:2009vd}
to be part of a family of odd-sphere blackfolds in a Minkowski background.
The thermodynamics of the odd-sphere (A)dS blackfolds will be obtained and
used in order to analyze the thermodynamic stability.

In parallel with \cite{Emparan:2009cs,Emparan:2009vd} we will also consider
the case of ellipsoidal even-ball blackfolds in (A)dS. For the AdS case, depending on the value of $\Omega L$ (with $L$ the (A)dS scale), the resulting blackfold solution can be related to either the ``ultra-spinning'' limit of
the higher-dimensional Kerr-AdS black hole found in \cite{Caldarelli:2008pz} or to a new ultra-spinning limit found in this paper. We also show that for the dS case only the latter limit is relevant. We then turn to a study of helical black rings in (A)dS,
which, like helical rings \cite{Emparan:2009vd} in Minkowski space,
provide evidence for the possibility of saturating the rigidity theorem
in (A)dS.

There are several motivations for our study. First of all, it is interesting to see if and how the new
families of asymptotically flat blackfold solutions found in \cite{Emparan:2009vd} generalize to different backgrounds, in this
case (A)dS. More generally it provides a wider arena in which to apply the blackfold method and compare to
exactly known neutral (A)dS black hole solutions, such as the higher-dimensional Kerr-(A)dS black holes \cite{Hawking:1998kw,Gibbons:2004uw} (henceforth
denoted by Kerr-(A)dS${}_D$). Also, as remarked above, constructing exact black hole spacetimes in (A)dS backgrounds is generally harder then in asymptotically flat spaces. In such cases, this makes the use of an alternate method,
like the blackfold approach, even more pertinent if one wishes to discover new horizon topologies.

Another point worth mentioning is that knowledge of possible black hole
phases can provide indications about  classical instabilities
and new inhomogeneous (`pinched') phases. The prototypical example of
this is the proposed connection \cite{Emparan:2007wm} (see also \cite{Emparan:2010sx})
between MP black holes \cite{Myers:1986un} and black rings through a merger transition involving  `pinched' black holes \cite{Emparan:2003sy} with spherical horizon topology. The emergence of this new stationary phase is related to the onset of the ultra-spinning Gregory-Laflamme (GL) instability of MP black holes found in Ref.~\cite{Emparan:2003sy}. Indeed, this has been verified in \cite{Dias:2009iu} by numerical analysis. As a consequence, analysis of (A)dS blackfolds
 can likewise be used to argue for possible mergers and connections
 between different black hole phases.

Finally black hole geometries in (A)dS backgrounds are interesting in view
of the AdS/CFT correspondence, notably the fluid/gravity correspondence \cite{Bhattacharyya:2008jc} which provides a fluid-dynamical description of AdS black holes.
The studies in  \cite{Bhattacharyya:2008ji} have,
however, not found evidence for new types of large AdS black objects. Interestingly though, plasma rings and plasma balls, including pinched plasma balls, have turned up in the study of hydrodynamic solutions in a theory with a confining vacuum \cite{Lahiri:2007ae}.
Such solutions correspond to large rotating black holes and black rings in
the dual Scherk-Schwarz compactified AdS${}_D$ space.
The analysis of \cite{Caldarelli:2008pz} and of the present paper
does not pertain to large AdS black objects, nor does it involve Scherk-Schwarz compactification. Nevertheless, it is perhaps possible to extend the results on AdS blackfolds to make a connection with these fluid-dynamical descriptions.

The outline of this paper is as follows. In Sec.~2 we give a general overview of the basic principles of the blackfold approach and set up the necessary formalism and notational conventions for what follows. In Sec.~3 we initiate our search for new horizon topologies in (A)dS spacetime, constructing a wide class of blackfolds whose horizon topology is the product of odd-spheres times a transverse sphere. These geometries generalize the thin (A)dS black rings of Ref.~\cite{Caldarelli:2008pz}
and are thermodynamically unstable in AdS, although they can be stable for certain values of the parameters in dS spacetime. In Sec.~4 we focus on
singly-spinning blackfolds with pancaked geometries, while in App.~\ref{as}  we deal with the multi-spinning case. We find that these even-ball (A)dS blackfolds
can be identified with two limits of the Kerr-(A)dS${}_D$ black hole. One of these limits was
previously found in \cite{Caldarelli:2008pz} while the other describes a new ultra-spinning limit of Kerr-(A)dS${}_D$ black holes in which the cosmological constant is taken to be small compared to the transverse section of the horizon (see App.~\ref{ab} for this limit in the general multi-spinning case). Our analysis predicts that  Kerr-(A)dS${}_D$ black holes in $D\geq 6$ suffer from an ultra-spinning GL-type instability when $\Omega_{i} L>1$. This is in agreement with  the recent numerical analysis of Ref.~\cite{Dias:2010gk} for the singly-spinning case in AdS.  In Sec.~5 we look for exotic topologies such as helical black rings in (A)dS. We find equilibrium conditions for these helical rings and, more generally, evidence of black hole solutions that saturate the rigidity theorem in any background with spherical symmetry. Finally, in Sec.~6, we summarize
and discuss the results found in this paper and give directions for future work.

\section{Brief tour on the blackfold method}
In this section we give a general overview of the blackfold approach and outline  the basics of the blackfold formalism developed in \cite{Emparan:2009cs,Emparan:2009at}, which the reader should consult if further clarification is needed. We also write down our conventions for the metric of the backgrounds under consideration.

\subsubsection*{Basic philosophy}
A \emph{blackfold} is essentially a thin black brane whose worldvolume spans a curved submanifold of a background spacetime, and the blackfold theory
developed in Refs.~\cite{Emparan:2009cs,Emparan:2009at}
 describes how to bend the worldvolume of
a black brane in a given background. The blackfold approach is an effective theory that captures the existence of black objects for which
the horizon is characterized by (at least) two widely separated length scales such that near any small enough region around the blackfold its geometry looks like that of a boosted flat black brane.
In particular, in the case of neutral uniform black branes, to which
we restrict ourselves in this paper,  the near-horizon metric is to leading order described by
\begin{equation} \label{E:dsb}
ds_{p-{\rm brane}}^2=\left(\gamma_{ab}+\frac{r_{0}^n}{r^n}u_{a}u_{b}\right)d\sigma^{a}d\sigma^{b}+\frac{dr^2}{1-\frac{r_{0}^n}{r^n}}+r^2d\Omega_{n+1}^2 \ ,
\end{equation}
where $\sigma^{a}$, $a=0 \ldots p$ are the worldvolume coordinates.
Thus the local adjustable parameters are the horizon thickness $r_{0}(\sigma^{a})$ and $p$ independent spatial components of the velocity field $u^{i} (\sigma^a)$, $i=1\ldots p$. The induced metric $\gamma_{ab}$
depends also on $\sigma$ via the embedding coordinates $X^\mu (\sigma^a)$
of the brane. The condition that the blackfold is locally a
flat black brane requires more formally that $r_0 \ll R $
where the scale $R$ is determined by the smallest intrinsic or extrinsic curvature radius of the worldvolume, i.e. the curvature of the worldvolume cannot be felt locally.

Since in this paper we will be concerned in applying these ideas to blackfolds
in $(A)dS_{D \geq 5}$ backgrounds, which naturally introduces
 its own length scale set by the cosmological constant $\Lambda$, we consider blackfolds in the regime
\begin{equation}
\label{validity}
r_{0}(\sigma^{a}) \ll {\rm min}(R,|\Lambda|^{-\frac{1}{2}}) \ ,
\end{equation}
so that neither the curvature of the worldvolume nor the curvature set by the (A)dS radius are felt locally at the blackfold.
This implies, in particular, that locally the blackfold is still described
by the asymptotically flat neutral black brane solution of \eqref{E:dsb}.  In the regime under consideration,
Ref.~\cite{Caldarelli:2008pz} considered the blackfold construction of thin rotating black rings in (A)dS space with horizon topology $S^1 \times S^{D-3}$.
Here, we will find more general classes of stationary solutions. In order to do so, we review the necessary formalism below.

\subsubsection*{Short review of the blackfold equations}

A \emph{blackfold} is specified by a set of collective coordinates $X^{\mu}(\sigma^{a})$ that describe the embedding of the $p+1$-dimensional worldvolume $\mathcal{W}_{p+1}$ of a black $p$-brane into a specific $D$-dimensional background spacetime, where the worldvolume $\mathcal{W}_{p+1}$ is spanned by the set of coordinates $\sigma^{a}$. Here the total spacetime dimension $D$ and $p$ are related as $D=n+p+3$, with $n \ge 1 $ being the number of spatial dimensions of the transverse space to the blackfold.
The effective stress tensor $T^{\mu\nu}$, supported on $\mathcal{W}_{p+1}$, encodes the degrees of freedom associated with the scale of the thickness of the $p$-brane, which are integrated out.
To leading order this  stress tensor is of perfect fluid form .
Conservation of the stress tensor is  equivalent to the system \cite{Emparan:2009at}
\begin{gather}
T^{\mu\nu}K_{\mu\nu}^{\rho}=0 ~~({\rm extrinsic~equations}) \ , \label{E:ext}\\
D_{a}T^{ab}=0 ~~({\rm intrinsic~equations}) \ ,  \label{E:bfe}
\end{gather}
where $D_{a}$ stands for the covariant derivative with respect to the induced metric on the worldvolume $\gamma_{ab}$, $T^{ab}$ is the world-volume energy-momentum tensor and $K_{\mu\nu}^{\rho}$ are the components of the extrinsic curvature tensor (second fundamental form).%
\footnote{We refer to App.~A of Ref.~\cite{Emparan:2009at} for the details on how to compute these quantities from a specific embedding $X^{\mu}(\sigma^{a})$.}

For stationary blackfolds, to which we restrict ourselves in this paper, the intrinsic equations are automatically satisfied by aligning the
fluid velocity field $u^\mu$ with a background Killing vector field $\textbf{k}$ that generates the isometries of the worldvolume $\mathcal{W}_{p+1}$. In general, if $\xi$ is a timelike Killing vector generator of the asymptotic time translations of the background and $\chi_{i}$ are spacelike ones, generators of asymptotic rotations, then $\textbf{k}$ takes the form
\begin{equation}
\textbf{k}=\xi+\sum_{i}\Omega_{i}\chi_{i} \ .
\end{equation}
One can then define the redshift factor $R_{0}$ between infinity and the blackfold worldvolume and the proper radii $R_{i}$ of the orbits generated by $\chi_{i}$ along the worldvolume as the norm of this set of commuting Killing vectors on the worldvolume:
\begin{equation} \label{E:r0r}
R_{0}=\sqrt{-\xi^2}|_{\mathcal{W}_{p+1}} \ , ~R_{i}=\sqrt{\chi_{i}^2}|_{\mathcal{W}_{p+1}} \ .
\end{equation}
It follows from this that $\textbf{k}$ can be expressed in a more convenient way
as
\begin{equation}
\label{ksol}
\textbf{k}=R_{0}\sqrt{1-V^2} \ ,
\end{equation}
where the velocity field $V$ is defined as
\begin{equation}  \label{E:vf}
V^{2}=\frac{1}{R_{0}^2}\sum_{i}\Omega_{i}^2R_{i}^2 \ .
\end{equation}
The horizon thickness $r_{0}$ can then be written as
\begin{equation}  \label{E:r0th}
r_{0}(\sigma^{a})=\frac{nR_{0}(\sigma^{a})}{2\kappa}\sqrt{1-V^2(\sigma^{a})} \ ,
\end{equation}
where $\kappa$ is the surface gravity.

\subsubsection*{Action and physical quantities}

Given the solution \eqref{ksol}-\eqref{E:r0th} of the intrinsic equations for stationary solutions, it was shown in Ref.~\cite{Emparan:2009at} that the extrinsic equations \eqref{E:ext} can be integrated to an action
\begin{equation} \label{E:actg}
I[X^{\mu}(\sigma_{a})]=\int_{\mathcal{W}_{p+1}}d^{p+1}\sigma\sqrt{-\gamma}|\textbf{k}|^{n}=\beta\int_{\mathcal{B}_{p}}dV_{(p)}R_{0}|\textbf{k}|^{n} \ ,
\end{equation}
where $\gamma$ is the determinant of the induced metric $\gamma_{ab}$, $\beta$ is an integration constant (that we will henceforth omit) and $\mathcal{B}_{p}$ is the spatial part of the blackfold worldvolume $\mathcal{W}_{p+1}$. The physical properties of the resulting blackfold solutions can then be easily computed. The total mass $M$ and angular momenta $J_i$ read
\begin{equation} \label{E:m}
M=\frac{\Omega_{(n+1)}}{16\pi G}\left(\frac{n}{2\kappa}\right)^{n}\int_{\mathcal{B}_{p}}dV_{(p)}R_{0}^{n+1}(1-V^2)^{\frac{n-2}{2}}(n+1-V^2) \ ,
\end{equation}
\begin{equation}
J_{i}=\frac{\Omega_{(n+1)}}{16\pi G}\left(\frac{n}{2\kappa}\right)^{n}n\Omega_{i}\int_{\mathcal{B}_{p}}dV_{(p)}R_{0}^{n-1}(1-V^2)^{\frac{n-2}{2}}R_{i}^2 \ ,
\end{equation}
while the entropy is given by
\begin{equation}
\label{entro:odd}
S=\frac{\Omega_{(n+1)}}{4 G}\left(\frac{n}{2\kappa}\right)^{n+1}\int_{\mathcal{B}_{p}}dV_{(p)}R_{0}^{n+1}(1-V^2)^{\frac{n}{2}} \ .
\end{equation}
Furthermore, the total integrated tension \cite{Emparan:2009vd} takes the form
\begin{equation} \label{E:t}
\mathcal{T}=\frac{\Omega_{(n+1)}}{16\pi G}\left(\frac{n}{2\kappa}\right)^{n}\int_{\mathcal{B}_{p}}dV_{(p)}R_{0}^{n+1}(1-V^2)^{\frac{n-2}{2}}(p-(n+p)V^2) \ .
\end{equation}
Using the explicit expressions \eqref{E:m}-\eqref{E:t} and also that $T=\frac{\kappa}{2\pi}$ one finds that these physical quantities satisfy the Smarr relation%
\footnote{This relation was first derived in \cite{Harmark:2004ch}
for  flat black branes of vacuum gravity in $D$ dimensions.}
\begin{equation} \label{smarr}
(D-3)M=(D-2)\left(\sum_{i}\Omega_{i}J_{i} + TS\right)+\mathcal{T} \ .
\end{equation}
For asymptotically flat black hole solutions of the vacuum Einstein equations the tension $\mathcal{T}$ \cite{Harmark:2004ch} must vanish (see \cite{Emparan:2009vd}), but this is generally not true in an (A)dS${}_{D}$ background. In fact, for the
thin black rings constructed in \cite{Caldarelli:2008pz} this is not the case, nor, as we will see, for any of the other A(dS) blackfold solutions found in this paper.

We furthermore note that all the solutions found in this paper and in \cite{Emparan:2009vd} not only obey the Smarr relation \eqref{smarr}, but we
also empirically observe the following relation for the Gibbs free energy
$G$
\begin{equation}
\label{Grel}
G=M-\sum_{i}\Omega_{i}J_{i}-TS=\frac{TS}{n} \ .
\end{equation}
The origin of this relation is explained in the recent paper \cite{upcoming2},
which also provides a more general derivation of the Smarr relation \eqref{smarr}.

\subsubsection*{(A)dS${}_{D}$ metric conventions}

In this work we will make use of the Anti-de Sitter metric written in terms of two different coordinate systems. We first write the metric for global AdS${}_{D}$ spacetime in the form
\begin{gather} \label{E:dsads1}
ds^{2}=-\mathcal{V}(r)dt^{2}+\frac{dr^{2}}{\mathcal{V}(r)} + r^{2}d\Omega^{2}_{D-2}
\spa 0\le r\le\infty \spa \mathcal{V}(r)=1+\frac{r^{2}}{L^{2}} \ .
\end{gather}
It will also be convenient to work with a metric that highlights the existent $U(1)$ symmetries of the background spacetime. This new metric can be obtained by introducing a new radial coordinate $\rho$ defined as
\begin{equation}
r=\frac{\rho}{1-\frac{\rho^2}{4L^2}} \ ,
\end{equation}
thus bringing the AdS${}_{D}$ metric \eqref{E:dsads1} into homogenous (spatially conformally flat) coordinates
\begin{gather} \label{E:dsads2}
ds^{2}=-F(\rho)dt^{2}+H(\rho)^{-1}(d\rho^2 + \rho^{2}d\Omega^{2}_{D-2})
\spa 0\le\rho\le 2L, \\ F(\rho)=\left(\frac{1+\frac{\rho^{2}}{4L^{2}}}{1-\frac{\rho^{2}}{4L^2}}
\right)^2 \spa H(\rho)=\left(1-\frac{\rho^{2}}{4L^2}\right)^2 \ .
\label{FandH}
\end{gather}
The AdS radius $L$ is related to the cosmological constant $\Lambda$ by
\begin{equation} \label{E:L}
\Lambda=\frac{(D-2)(D-1)}{L^2} \ ,
\end{equation}
and thus the range of validity \eqref{validity} of the results in this paper can be recast as $r_{0}\ll {\rm min}(R,L)$. The dS${}_{D}$ metric in both coordinate systems can be obtained by performing a Wick rotation such that $L\to iL$ in the metrics \eqref{E:dsads1} and \eqref{E:dsads2}.

\section{Blackfolds with odd-sphere horizon topology} \label{S:odd}

In  Ref.~\cite{Emparan:2009vd} the blackfold approach was used to
construct a class of novel
 black holes in $D$-dimensional flat spacetime with horizon topology
\begin{equation}
\left(\Pi_{p_{a}={\rm odd}}S^{p_{a}}\right)\times s^{n+1} \spa \sum_{a=1}^{l}p_{a}=p \ .
\end{equation}
This class contains not only the family of thin black rings with horizon topology $S^{1}\times s^{n+1}$ but also single (and the product of) odd-spheres with $S^{2k+1}$ horizon geometry. In this section we generalize these results to (A)dS${}_{D}$ spacetime, and furthermore study the thermodynamic stability of these new solutions.

\subsection{Black $S^{2k+1}$-folds in AdS${}_{D}$} \label{S:s}

The first step for constructing a stationary blackfold solution is to embed the spatial worldvolume $\mathcal{B}_{p}$, $p=2k+1$  into the background space. In this case we want
to wrap the spatial world-volume on a $S^{2k+1}$ sphere embedded into a $(2k+2)$-dimensional spatially conformally flat subspace of AdS${}_{D}$ spacetime \eqref{E:dsads2}. The appropriate
 part of the background metric can be conveniently expressed as
\begin{gather} \label{E:dss1}
ds^2_{2k+2} = H(\rho)^{-1}\left(d\rho^2 + \rho^{2}\sum_{i=1}^{k+1}(d\mu_{i}^2+\mu_{i}^2d\phi_{i}^2)\right)
\spa \sum_{\mu = 1}^{k+1}\mu_{i}^2=1\ ,
\end{gather}
so that the $S^{2k+1}$ is parameterized by $k+1$ Cartan angles $\phi_{i}$ and $k$ independent director cosines $\mu_{i}$. It is then natural to choose a gauge in which the worldvolume $\mathcal{B}_{2k+1}$ is specified by the embedding scalar $\rho=\bar{R} (\{ \mu_i \})$ and the spatial worldvolume coordinates
\begin{equation}
\{\mu_{i},~i=1,...,k\},~\{\phi_{i},~i=1,...,k+1\} \ .
\end{equation}
In order to construct the action for these blackfolds one needs the induced metric on the worldvolume. In terms of the Cartan angles and director cosines this metric takes the form
\begin{equation}
\begin{split}
ds^{2}_{2k+1}&=H(\bar{R}(\mu_{i}))^{-1} \sum_{i,j=1}^{k}\left( \left(\delta_{ij}+\frac{\mu_{i}\mu_{j}}{\mu_{k+1}^2}\right)\bar{R}(\mu_{i})^2 +\partial_{i}\bar{R}(\mu_{i})\partial_{j}\bar{R}(\mu_{j})\right) d\mu_{i}d\mu_{j}\\
&+H(\bar{R}(\mu_{i}))^{-1}\bar{R}(\mu_{i})^2\sum_{i=1}^{k+1}\mu_{i}^2d\phi_{i}^2 \ .
\end{split}
\end{equation}
Since, in order to have a stationary blackfold, the corresponding Killing vector must generate isometries of the worldvolume, the horizon Killing vector takes the form
\begin{equation}
\textbf{k}=\frac{\partial}{\partial t}+\sum_{i=1}^{k+1}\Omega_{i}\frac{\partial}{\partial\phi_{i}} \ .
\end{equation}

The redshift factor $R_{0}$ and the proper radii $R_{i}$ of the orbits generated by $\frac{\partial}{\partial\phi_{i}}$ are given respectively by $R_{0}=\sqrt{F(\bar{R}(\mu_{i}))}$ and $R_{i}=H(\bar{R}(\mu_{i}))^{-\frac{1}{2}}\bar{R}(\mu_{i})$, while the velocity field \eqref{E:vf} becomes
\begin{equation}
 V(\mu_{i})^2=\frac{\bar{R}(\mu_{i})^2}{(1+\frac{\bar{R}(\mu_{i})^2}{4L^2})^2}\sum_{i=1}^{k+1}\mu_{i}^2\Omega_{i}^2 \ .
\end{equation}
 We recall
that the functions $F$ and $H$ entering the background metric
are defined in \eqref{FandH}.

For simplicity we restrict to round odd-spheres, so that we take
the scalar $\bar{R}$ to be constant.
Furthermore, we are interested in the maximally symmetric case for which the $S^{2k+1}$ sphere is rotating with equal angular velocity $\Omega$ in all $k+1$ directions $\phi_{i}$
It follows that  the action \eqref{E:actg} reduces to an $\bar{R}$-dependent potential of the form
\begin{equation}\label{E:acts1}
I[\bar{R}]= \Omega_{(p)}\sqrt{F(\bar{R})}H(\bar{R})^{-\frac{p}{2}}\bar{R}^{p}(F(\bar{R})-\Omega^2H(\bar{R})^{-1}\bar{R}^2)^{\frac{n}{2}} \ ,
\end{equation}
where $p=2k+1$ and $\Omega_{(p)}$ is the area of the $S^{2k+1}$ sphere. A nicer form of the action can be obtained by performing the inverse transformation between the coordinate systems \eqref{E:dsads1} and \eqref{E:dsads2}.
Thus defining $R=\bar{R}/(1-\frac{\bar{R}^2}{4L^2})$, the action \eqref{E:acts1} above becomes\footnote{In fact, in this highly symmetrical case we could have simply used the form of the metric \eqref{E:dsads1} and obtained this action straight away.}
\begin{equation}\label{E:acts2}
I[R]= \Omega_{(p)}R_{0}R^{p}(R_{0}^2-\Omega^2R^2)^{\frac{n}{2}} \ ,
\end{equation}
where now $R_{0}=\sqrt{\mathcal{V}(R)}$, with $\mathcal{V}$ defined in \eqref{E:dsads1}.
 Varying this action with respect to $R$ we obtain the equilibrium condition for $\Omega$
\begin{equation} \label{E:oms}
\Omega^2=\frac{1 + \textbf{R}^2 }{R^2}\frac{p+\textbf{R}^2(n+p+1)}{(n+p)+\textbf{R}^2(n+p+1)} \ ,
\end{equation}
where we have defined the dimensionless parameter $\textbf{R}=\frac{R}{L}$.
It is straightforward to check that the limit $L\to\infty$ gives the result obtained in \cite{Emparan:2009vd} for  $S^{2k+1}$-folds constructed in a Minkowski background and that the special case of a black ring in AdS${}_{D}$ $(p=1)$ agrees with the one obtained in \cite{Caldarelli:2008pz}. It should also be noted that the inverse of the relation \eqref{E:oms} above in terms of $R$ is single valued for a fixed value of $\Omega$ and is valid for all values of $L$.

\subsubsection*{Physical properties and thermodynamic stability}\label{S:sp}

The physical properties for the odd-sphere AdS blackfolds are straightforwardly obtained using equations \eqref{E:m}-\eqref{E:t}. Setting $V_{(p)}=R^{p}\Omega_{(p)}$ for the volume of the $S^{2k+1}$ sphere we find
\begin{equation} \label{E:masss}
M=\frac{\Omega_{(n+1)}V_{(p)}}{16\pi G}r_{0}^{n}(1+\textbf{R}^2)^{\frac{3}{2}}(1+n+p) \ ,
\end{equation}
\begin{equation} \label{E:ens}
S=\frac{\Omega_{(n+1)}V_{(p)}}{4G}r_{0}^{n+1}\sqrt{\frac{\textbf{R}^2+(n+p)(1+\textbf{R}^2)}{n}} \ ,
~ T=\frac{n}{4\pi r_{0}}\sqrt{\frac{n \left(1+\textbf{R}^2\right)}{\left(1+\textbf{R}^2\right) (n+p)+\textbf{R}^2}} \ ,
\end{equation}
\begin{equation} \label{E:ans}
J_{i}=\frac{2}{p+1}\frac{\Omega_{(n+1)}V_{(p)}}{16\pi G}r_{0}^{n} R \sqrt{\left(p+\textbf{R} (n+p+1)\right) \left((n+p)+\textbf{R}^2(n+p+1)\right)} \ ,
\end{equation}
\begin{equation} \label{E:oms2}
\Omega_{i}=\Omega \spa i=1,...,k+1\ .
\end{equation}
Moreover, the total tension $\mathcal{T}$ becomes
\begin{equation} \label{E:tens}
\mathcal{T}=-\frac{\Omega_{(n+1)}V_{(p)}}{16\pi G}r_{0}^{n}(1+\textbf{R}^2)^{\frac{1}{2}}\textbf{R}^2 (n+p+1) \ ,
\end{equation}
showing explicitly that in AdS spacetime this quantity is not necessarily zero.
The physical quantities above can be shown to satisfy the Smarr relation \eqref{smarr} and the relation \eqref{Grel}.

In order to study the thermodynamic stability of these blackfolds one needs to compute the specific heat at constant angular momenta $C_{J}$ and the spectrum of the isothermal differential moment of inertia tensor $\epsilon^{ij}$, given by
\begin{equation}
C_{J}=T\left(\frac{\partial S}{\partial T}\right)_{J} \spa \epsilon^{ij}=\left(\frac{\partial J^{i}}{\partial \Omega_{j}}\right)_{T} \ .
\end{equation}
In the present case the tensor $\epsilon^{ij}$ is diagonal and has all its diagonal elements equal to each other, so  $\epsilon^{ii}=\epsilon$. The condition for thermodynamic stability reads (see e.g. \cite{Monteiro:2009tc})
\begin{equation}
C_{J}>0 \spa {\rm spec}(\epsilon^{ij})>0 \ ,
\end{equation}
where ``spec`` denotes the spectrum of eigenvalues.
We will now proceed by computing these quantities for two different cases:

\noindent \textbf{\emph{(i)}} Flat case ($L\to\infty$ limit)
\begin{equation}
C_{J}=-\frac{\Omega_{(n+1)} V_{(p)}r_{0}^{n+1}}{4G} \frac{\sqrt{\frac{n+p}{n}} (n+p+1)}{(p+1)} \ ,
\end{equation}
\begin{equation}
\epsilon=-\frac{\Omega_{(n+1)} V_{(p)} r_{0}^n(p+1)  (n+p) R^{2}}{16 \pi  G} \ .
\end{equation}
Since both $C_{J}<0$ and $\epsilon<0$ we conclude that $S^{2k+1}$-folds in a Minkowski background are thermodynamically unstable. This is expected since asymptotically flat black hole solutions generally show this type of behavior.
\\ \\
\textbf{\emph{(ii)}} AdS case (finite $L$)
\begin{equation}
C_{J}=-\frac{\Omega_{(n+1)} V_{(p)}r_{0}^{n+1}C_{1}}{C_{2}}(1+\textbf{R}^2) (n+p+1) \sqrt{\frac{\left(1+\textbf{R}^2\right) (n+p)+\textbf{R}^2}{n}} \ ,
\end{equation}
with\\
$C_{1}=( p (n+p)+2 \textbf{R}^2 (n^2+2 n p+n+p (p+2)) +\textbf{R}^4 (n+p+1) (2 n+p+3))$ \ , \\
$C_{2}= 4 G \left(p (p+1) (n+p)+\textbf{R}^2 (n+p+1) (n (p+2)+p (3 p+4))+ C_{3}\right)$ \ ,  \\
$C_{3}=\textbf{R}^4 (n+p+1) (2 n (p+2)+p (3 p+8)+3)+(p+3) \textbf{R}^6(n+p+1)^2$ \ , \\
and
\begin{equation}
\epsilon= -\frac{\Omega_{(n+1)}V_{(p)}r_{0}^nD_{1}}{D_{2}}\sqrt{1+\textbf{R}^2}  R^{2} \sqrt{\frac{p+\textbf{R}^2 (n+p+1)}{\left(1+\textbf{R}^2\right) (n+p)+\textbf{R}^2}}(\textbf{R}^2+(n+p)(1+\textbf{R}^2) ) \ ,
\end{equation}
with,\\
$D_{1}=\left( p (p+1) (n+p)+\textbf{R}^2 (n+p+1) (n (p+2)+p (3 p+4))+ D_{3}\right)$ \ , \\
$D_{2}=16 \pi  G \left(1+\textbf{R}^2\right) \left(p (n+p)+2 p \textbf{R}^2 (n+p+1)+ (1+p)(1+n+p)(\textbf{R}^4\right)$ \ , \\
$D_{3}=\textbf{R}^4 (n+p+1)(2 n (p+2)+p (3 p+8)+3)+(p+3)\textbf{R}^6 (n+p+1)^2$ \ . \\ \\
By taking a closer look at $C_{2}$ we see that since $L>0$ the denominator is always positive while all other terms $C_{1},C_{3}$ are positive, hence $C_{J}$ is always negative due to the negative overall minus sign. Similarly, by looking at the expression for $\epsilon$ we conclude that all terms $D_{1},D_{2},D_{3}$ are positive and hence $\epsilon$ is negative for all values of $L$. Therefore these blackfolds are thermodynamically unstable. This is also to be expected since small black holes in AdS spacetime are known to be unstable.

\subsection{The general product of odd-spheres in AdS${}_{D}$ \label{sec:prod}}

The class discussed above is part of a larger one in which in which the spatial worldvolume $\mathcal{B}_{p}$ is a product of $l$ round odd-spheres embedded as in Sec.~\ref{S:s} above. We label the different spheres by an index $a=1,...,l$ and denote $\bar{R}_{a}$ as the corresponding (constant) radius of each $S^{p_{a}}$, where $p_{a}$ is an odd integer. For the sake of simplicity we choose for each sphere the angular velocity associated with each Cartan angle direction  to be equal
\begin{equation}
\Omega_{i}^{(a)}=\Omega^{(a)} \spa \forall i=1,...,\frac{p_{a}+1}{2} \ .
\end{equation}
To embed $\mathcal{B}_{p}$ we consider a conformally flat $(p+l)$-dimensional subspace of AdS${}_{D}$ with the metric
\begin{equation}
ds^{2}_{p+l}=H(\rho)^{-1}\sum_{a=1}^{l}\left(d\rho_{a}^2+\rho_{a}^2d\Omega_{p_{a}}^2\right)
\spa \rho^2=\sum_{a=1}^{l}\rho_{a}^2 \spa \sum_{a=1}^{l}p_{a}=p \ .
\end{equation}
Again we choose the Cartan angles and director cosines of each $S^{p_{a}}$ sphere as the spatial worldvolume coordinates and take $\rho_{a}=\bar{R}_{a}$ as the embedding scalars. The transverse space is $(n+2-l)$-dimensional, hence we require that $l\le n+2$.

Defining $\bar{R}^2=\sum_{a=1}^{l}\bar{R}_{a}^2$, the action \eqref{E:acts1} can be generalized to an $\bar{R}_{a}$-dependent potential
\begin{equation} \label{E:actsp11}
I[\{\bar{R}\}]= \Pi_{b=1}^{l}\Omega_{(p_{b})}\sqrt{F(\bar{R})}\bar{R}_{b}^{p_{b}}H(\bar{R})^{-\frac{p_{b}}{2}}\left(F(\bar{R})-H(\bar{R})^{-1}\sum_{a=1}^{l}(\Omega^{(a)}\bar{R}_{a})^{2}\right)^{\frac{n}{2}} \ .
\end{equation}
Introducing new scalars $R_{a}$ as $R_{a}=\bar{R}_{a}/(1-\frac{\bar{R}_{a}^2}{4L^2})$ the previous action can be put in a simpler form
\begin{equation} \label{E:actsp1}
I[\{R\}]= \Pi_{b=1}^{l}\Omega_{(p_{b})}R_{0}R_{b}^{p_{b}}\left(R_{0}^2-\sum_{a=1}^{l}(\Omega^{(a)}R_{a})^{2}\right)^{\frac{n}{2}} \ ,
\end{equation}
where $R_{0}=1-\frac{R^2}{L^2}$, with $R^2=\sum_{a=1}^{l}R_{a}^2$. Varying this action with respect to each of the scalars $R_{a}$ gives rise to $l$ coupled equations for each of the angular velocities $\Omega^{(a)}$. The equilibrium condition can then be found to be
\begin{equation}
\left(\Omega^{(a)}\right)^2=\frac{1+\textbf{R}^2}{R_{a}^2}\frac{ p_{a}+\textbf{R}_{a}^2 (n+p+1)}{
   (n+p)+ \textbf{R}^{2}(n+p+1)} \ ,
 \end{equation}
where we have defined $\textbf{R}_{a}=R_{a}/L$. It  easy  to check that in the limit $L\to\infty$ the above condition agrees with that of \cite{Emparan:2009vd} while the particular case $l=1$ agrees with \eqref{E:oms}.

The physical properties for these blackfolds are also easily computed. In fact, the expressions for $M,~S,~T,~\mathcal{T}$ coincide with those in \eqref{E:masss}, \eqref{E:ens}, \eqref{E:tens} for  a single odd-sphere if we define the volume of $\mathcal{B}_{p}$ as $V_{(p)}=\Pi_{a}V_{(p_{a})}$ while the angular momenta and angular velocities read
\begin{equation}
J^{(a)}_{i}=\frac{2}{p_{a}+1}\frac{\Omega_{(n+1)}V_{(p)}}{16\pi G}r_{0}^{n}R_{a}\sqrt{(p_{a}+\textbf{R}_{a}^2(n+p+1))((n+p)+\textbf{R}^{2}(n+p+1))} ~.
\end{equation}
The Smarr relation \eqref{smarr} and the relation \eqref{Grel} can also be verified  for this case.

\subsection{Black $S^{2k+1}$-folds in dS${}_{D}$}

The equilibrium condition for odd-sphere blackfolds in a de Sitter background $dS_D$ can be easily obtained from those in \eqref{E:oms} by performing the Wick rotation $L\to i L$, leading to
\begin{equation} \label{E:omsd}
\Omega^2=\frac{1-\textbf{R}^2}{R^2}\frac{\textbf{R}^2(n+p+1)- p}{\textbf{R}^2(n+p+1)-(n+p)} \ .
\end{equation}
Since $\Omega^2$ might become negative for certain values of the parameters we must impose the condition $\Omega^2\ge0$, which implies that the ratio $\textbf{R}$ should be constrained to the region
\begin{equation} \label{E:S2komegadSc}
\sqrt{\frac{n+p}{n+p+1}}\le \mathbf{R}\le 1~\vee~\sqrt{\frac{p}{n+p+1}}<\mathbf{R} \ .
\end{equation}
Hence black $S^{2k+1}$-folds in  $dS_D$  do not exist for all values of $\mathbf{R}$. Moreover, a static solution always exists if\footnote{The case for which $\textbf{R}=1$ leads to a blackfold with vanishing horizon and vanishing physical properties.}
\begin{equation} \label{E:conds}
\textbf{R}^2=\frac{p}{n+p+1} \ .
\end{equation}

The physical properties of these solutions can be obtained from those given in Sec.~\ref{S:sp} by taking into account the same Wick rotation. A general analysis of the thermodynamic stability of these blackfolds for all values of the parameters is altogether cumbersome, therefore we analyze in detail two particular cases.

\noindent \textbf{\emph{(i)}} Black Ring in dS$_{5}$ ($n=1$, $p=1$)\newline
In this case an equilibrium value for $\Omega$ exists if
\begin{equation}
\sqrt{\frac{2}{3}}\le \textbf{R}\le 1~\vee~\frac{1}{\sqrt{3}}<\textbf{R} \ .
\end{equation}
Then the specific heat $C_{J}$ and the spectrum of $\epsilon^{ij}$ can be computed to be
\begin{equation}
C_{J}=\Omega_{(2)}V_{(1)}\frac{3 r_{0}^2\sqrt{\frac{1}{2-3\textbf{R}^2}}(-1+\textbf{R}^2)(-7\textbf{R}^2+9\textbf{R}^4)}{4 G  (1-6(\textbf{R}^2+\textbf{R}^4))} \ ,
\end{equation}
\begin{equation}
\epsilon=\Omega_{(2)}V_{(1)}\sqrt{1-\textbf{R}^2}\frac{r_{0}(2R-\textbf{R}^3)^{3}(1-6\textbf{R}^2+6\textbf{R}^4)}{16 G \pi\sqrt{2-9\textbf{R}^2+9\textbf{R}^4} (1-4\textbf{R}^2+6\textbf{R}^4-3\textbf{R}^6)} \ .
\end{equation}
A detailed analysis leads us to the conclusion that black rings in $dS_{5}$ spacetime are thermodynamically stable if
\begin{equation}
\frac{2}{1+\sqrt{13}}\le \textbf{R}\le\frac{1}{\sqrt{3+\sqrt{3}}} \ .
\end{equation} \\
\textbf{\emph{(ii)}} Black $S^{3}$-fold in dS$_{7}$ ($n=1$, $p=3$) \newline
An equilibrium value for $\Omega$ can be found to exist if
\begin{equation}
 \sqrt{\frac{4}{5}}\le \textbf{R}\le1~\vee~\sqrt{\frac{3}{5}}<\textbf{R} \ .
\end{equation}
The specific heat and the spectrum of $\epsilon^{ij}$ are given by
\begin{equation}
C_{J}=\frac{\Omega_{(2)}V_{(3)}5r_{0}^2 \sqrt{\frac{1}{4 -5 \textbf{R}^2}} \left(\textbf{R}^2-1\right) \left(6-23 \textbf{R}^2+20 \textbf{R}^4\right)}{4 G \left(6-20 \textbf{R}^2+15 \textbf{R}^4\right)} \ ,
\end{equation}
\begin{equation}
\epsilon=\frac{\Omega_{(2)}V_{(3)}R^2r_{0}\left(4 -5 \textbf{R}^2\right)^3 \sqrt{1-\textbf{R}^2} \left(6-20 \textbf{R}^2+15 \textbf{R}^4\right) \sqrt{\frac{1}{5 \textbf{R}^2-4 }+1}}{16 \pi  G
   (1-\textbf{R}^2) \left(6-15 \textbf{R}^2+10 \textbf{R}^4\right) \sqrt{12-35 \textbf{R}^2+25 \textbf{R}^4}} \ .
\end{equation}
Hence black $S^{3}$-folds in $dS_{7}$ are thermodynamically stable only if
\begin{equation}
 \sqrt{\frac{2}{5}}\le \textbf{R}\le\sqrt{\frac{6}{10-\sqrt{10}}} \ .
\end{equation}

It would be interesting to examine the form of the stability criteria
for the general case and understand the physical origin of the range of
stability observed for these odd-sphere dS blackfolds.

\section{Ultra-spinning and "ultra-spinning" Kerr-(A)dS${}_{D}$ black holes as blackfolds} \label{S:even}

Blackfold solutions in flat space with $\mathcal{B}_{p}$ an even-dimensional ellipsoidal ball have been shown to exist
in Refs.~\cite{Emparan:2009cs,Emparan:2009vd}. These have
event horizon with $S^{D-2}$ topology due to the fact that the
transverse $S^{n+1}$ is non-trivially fibered over the ellipsoid, becoming zero size at the boundary.  In fact, the physical properties of these even-ball blackfolds  have been shown to exactly reproduce those of ultra-spinning MP black holes \cite{Emparan:2003sy}.

Despite the fact that ultra-spinning regimes have not been found for spinning black holes in AdS${}_{D}$, it has been pointed out in \cite{Caldarelli:2008pz} that the Kerr-AdS${}_{D}$ black hole with an appropriate choice of mass and rotation parameters $m,a$ has an "ultra-spinning" regime which shares many of the same properties with the ultra-spinning regime of the MP black hole. As an example, the transverse and parallel size of the horizon of the single spinning Kerr-AdS${}_{D}$ black hole in $D\geq 6$ behave in the limit $a\to L$ as
\begin{equation} \label{E:div}
l_{\perp}\sim r_{+} \spa l_{\parallel}\sim\left(\frac{r_{+}^2+a^2}{\Xi}\right)^{\frac{1}{2}} \spa
\Xi \equiv 1-\frac{a^2}{L^2} \ ,
\end{equation}
where $r_{+}$ is the event horizon radius. For fixed mass the ratio $\frac{l_{\parallel}}{l_{\perp}}$ diverges like $\sim\Xi^{-\frac{D-1}{2(D-5)}}$, meaning that the horizon pancakes out along the plane of rotation. Thus this limit could in principle be captured by an even-dimensional ellipsoidal ball $\mathcal{B}_{p}$,
in particular a disk in the case of one plane of rotation. Moreover, the Kerr-AdS${}_{D}$ solution has a BPS bound $J\le L M$ \cite{Chrusciel:2006zs} restricting the rotation parameter in such a way that $a\le L$. Thus it is clear that at fixed mass $M$ and fixed $L$ one cannot simply take $a\to\infty$ and obtain an ultra-spinning limit as in the asymptotically flat case since the bound would be violated. However, as we will show below, it is possible to take a limit in which $a \rightarrow \infty$  and simultaneously taking $L\to\infty$ while keeping the ratio $\frac{a}{L}$ constant.
This limit amounts to considering a very thin black hole compared to the scale $L$, set by the cosmological constant, while keeping $a$ of the same order of magnitude as $L$, $i.e.$, making the black hole simultaneously thin compared to the parallel section of the horizon. Thus the resulting limit is not asymptotically flat. Furthermore,
we will  show that this limit can also be captured by the same blackfold $\mathcal{B}_{p}$.

In this section we will start by solving the action for a worldvolume with even-dimensional ball geometry in an AdS${}_{D}$ background and compute the physical properties of these solutions.
Subsequently we will identify the properties of this solution
with both  the ultra-spinning and "ultra-spinning" regimes of the Kerr-(A)dS${}_{D}$ black hole. At the end of this section we will generalize these results to a dS${}_{D}$ background.

\subsection{Even-ball blackfolds in AdS${}_{D}$ } \label{S:eads}

The starting point for constructing these blackfolds is to consider a planar $2k$-fold embedded into a $(2k+1)$-dimensional spatially conformally flat subspace of AdS${}_{D}$, which can be equipped with the metric
\begin{equation}
ds^2_{2k+1} = H(\rho)^{-1}\left(dz^2+\sum_{i=1}^{k}(d\rho_{i}^2+\rho_{i}^2d\phi_{i}^2)+\sum_{j=1}^{D-2(k+1)}dx_{j}^2\right)
\spa \rho^2=\sum_{i=1}^{k}\rho_{i}^2+\sum_{j=1}^{D-2(k+1)}x_{j}^2 \ .
\end{equation}
It is then natural to choose the embedding of $\mathcal{B}_{2k}$ as
\begin{equation}
z=Z(\rho_{i}),~x_{j}=0,~j=1,...,D-2(k+1),~\{\rho_{i}=\sigma_{i},~\phi_{i}=\sigma_{i+1},~i=1,...,k\} \ ,
\end{equation}
with the function  $Z(\rho_{i})$ to be determined.
The Killing vector field that generates the isometries of the worldvolume
is of the form
\begin{equation}
\textbf{k}=\frac{\partial}{\partial t}+\sum_{i}^{k}\Omega_{i}\frac{\partial}{\partial\phi_{i}} \ .
\end{equation}
Thus the action \eqref{E:actg} takes the simple form
\begin{equation} \label{E:acta}
I[Z(\rho_{i})]=\Omega_{(p)} \int \prod_{i=1}^k d \rho_i R_{0}\prod_{j=1}^{k}R_{i}\sqrt{1+\partial_{\rho_{i}}Z(\rho_{i})}\left(R_{0}^2-\sum_{i=1}^{k}R_{i}^2\Omega_{i}^{2}\right)^{\frac{n}{2}} \ ,
\end{equation}
where $R_{0}=\sqrt{F(\rho)}$ and $R_{i}=\sqrt{H(\rho)^{-1}}\rho_{i}$. Varying this action with respect to $Z(\rho_{i})$ and analyzing the resulting equation leads to the conclusion that $Z=0$ is a blackfold solution. In the asymptotically flat case \cite{Emparan:2009vd}, any plane $Z={\rm const.}$ is a valid solution, while in the present case due to the AdS${}_{D}$ potential only $Z=0$ is a solution.

In what follows we will focus on the case of singly-spinning blackfolds ($k=1$) with angular momentum along the $\phi_{1}$ direction and deal with the general case in App.~\ref{as}. In this case the worldvolume velocity field is given by
\begin{equation}
V (\rho) =\frac{\rho}{1+\frac{\rho^2}{4L^2}}\Omega \ ,
\end{equation}
where $R=R_{1}$, $\Omega=\Omega_{1}$ and $\rho=\rho_{1}$. Since $V$ cannot exceed the
speed of light $V=1$, we find that $\rho$ is bounded by the maximum
value
\begin{equation} \label{E:cons1}
\rho_{\rm max}=2L(L\Omega - \sqrt{L^2\Omega^2-1}) \ .
\end{equation}
The other value for $\rho$ at which $V=1$, which has a plus sign in front of
the square root, can be discarded since in this coordinate system (see \eqref{E:dsads2}) spatial infinity is reached at $\rho=2L$.

Since the argument of the square root in \eqref{E:cons1} must be positive definite we obtain a constraint
\begin{equation} \label{E:const2}
0\le\alpha\le 1 \spa \alpha \equiv (L^2\Omega^2)^{-1} \ .
\end{equation}
In terms of the parameter $\alpha$ defined above we can now distinguish three different situations:\\
\textbf{\emph{(i)}} $\alpha=0$. In this case $\rho_{\rm max}\to\frac{1}{\Omega}$. This is the asymptotically flat space case and so we correctly recover the ultra-spinning MP black hole where $\rho_{\rm max}\sim a$, so that the blackfold has the shape of a disc with radius $a$ (see \cite{Emparan:2009cs,Emparan:2009vd}).\\
\textbf{\emph{(ii)}} $\alpha=1$. In this case $\rho_{max}\to 2L$, so  the disc extends all the way to spatial infinity. As we show below, this corresponds to the "ultra-spinning" limit taken in \cite{Caldarelli:2008pz} where $a\to L$. \\
\textbf{\emph{(iii)}} $0<\alpha<1$. In this case $\rho_{\rm max}<2L$, so
 that the disc is cut at some value of $\rho$ and does not reach spatial infinity.  As we will see below, this corresponds to a new ultra-spinning limit of the Kerr-AdS${}_{D}$ black hole.

Before taking these limits it is useful to compute the physical properties of this blackfold. These are easily obtained using the equations \eqref{E:m}-\eqref{E:t}, yielding
\begin{equation}\label{E:ulm}
M=\frac{\Omega_{(D-2)} }{8\pi G}\frac{\hat{r}_{+}^n}{(1-\alpha)^2}\left(1+\frac{(n+1)(1-\alpha)}{2}\right)\frac{1}{\Omega^2} \ ,
\end{equation}
\begin{equation}
J=\frac{\Omega_{(D-2)} }{8\pi G}\frac{\hat{r}_{+}^n}{(1-\alpha)^2}\frac{1}{\Omega^3} \spa
S=\frac{\Omega_{(D-2)} }{4G}\frac{\hat{r}_{+}^{n+1}}{(1-\alpha)}\frac{1}{\Omega^2} \ ,
\end{equation}
\begin{equation} \label{E:ten}
\mathcal{T}=-\alpha\frac{\Omega_{(D-2)} }{8\pi G}\frac{\hat{r}_{+}^{n}}{(1-\alpha)^2}\frac{1}{\Omega^2} \ ,
\end{equation}
where we have defined $\hat{r}_{+}=\frac{n}{2\kappa}$. It is worthwhile to notice that the tension $\mathcal{T}$ vanishes only if $\alpha=0$, in agreement with the flat space result. Also, when $\alpha$ lies within the region $0<\alpha\le1$ the tension is non-zero and hence the blackfold does not describe an asymptotically flat solution. Moreover, it is straightforward to check that the quantities above satisfy the Smarr relation \eqref{smarr} and the relation \eqref{Grel}.

We also note that by defining $r=\rho/(1-\frac{\rho^2}{4L^2})$ the thickness $r_{0}$ becomes
\begin{equation}\label{E:ut}
r_{0}=\frac{n}{2\kappa}\sqrt{1-r^2\Omega^2(1-\alpha)} \ ,
\end{equation}
so that in terms of this coordinate we now have $r_{\rm max}=\rho_{\rm max}/(1-\frac{\rho_{\rm max}^2}{4L^2})$. The thickness remains  finite for all values of $\alpha$ since when $\alpha=1$ and $r_{max}\to\infty$, $R_{0}\to\infty$ but $r_{0}\to0$. Thus the  blackfold
 is always in the regime $r_{0}\ll L$.

We will now proceed to identify the physical properties of the disc blackfold given above with those corresponding to the two different limits of the Kerr-AdS${}_{D}$ black hole.

\subsubsection*{$\alpha=1$: the "ultra-spinning" limit}

This limit was found in Ref.~\cite{Caldarelli:2008pz} and amounts to taking $a\to L$, and hence $\Omega\to\frac{1}{L},\alpha\to1$ while keeping $\hat{\mu}=\frac{2m}{L^2(1-\alpha)^2}$ finite, i.e., sending $m\to0$. The resulting metric near the rotation axis can be expressed in appropriately rescaled coordinates as
\begin{equation} \label{E:ultra1}
ds^2=\Xi^{\frac{4}{D-5}}\left(-\left(1-\frac{\hat{\mu}}{\hat{r}^{D-5}}\right)d\hat{t}^2+\left(1-\frac{\hat{\mu}}{\hat{r}^{D-5}}\right)^{-1}d\hat{r}^2 +\hat{r}^2d\Omega_{D-4}^2+d\sigma^{2}+\sigma^{2}d\phi^2\right) \ ,
\end{equation}
where $\Xi$ is given in \eqref{E:div}.
This metric describes the geometry of a flat black membrane with an overall conformal factor. Its physical properties can be summarized as follows
\begin{equation} \label{E:um1}
M=\frac{\Omega_{(D-2)} }{8\pi G}\hat{\mu}L^2 \ ,
\end{equation}
\begin{equation}
S=\frac{\Omega_{(D-2)} }{4G}r_{+}^{D-4}\frac{r_{+}^{2}+L^2}{(1-\alpha)} \spa T=\frac{D-5}{4\pi r_{+}} \ ,
\end{equation}
\begin{equation}\label{E:uj1}
J=\frac{\Omega_{(D-2)} }{8\pi G}\hat{\mu}L^3 \spa \Omega=\frac{1}{L} \spa
\end{equation}
\begin{equation} \label{E:iden1}
r_{+}=\left(\frac{2m}{L^2}\right)^{\frac{1}{(D-5)}} \ .
\end{equation}
It is easy to check that with the identification $\hat{r}_{+}=r_{+}$ and using $\Omega=L^{-1}$ the blackfold physical properties  \eqref{E:ulm}-\eqref{E:ten} found above exactly reproduce the properties \eqref{E:um1}-\eqref{E:uj1} of this "ultra-spinning" limit (note that $n=D-5$).
To see this one also needs to use the fact  that our blackfold is a valid solution only in the regime $L \gg r_{+}$, $i.e.$, the entropy becomes\footnote{In fact, as $a\to L$, the horizon size $r_{+}$ approaches zero and hence $S\to0$. The tension $\mathcal{T}$ remains finite in this limit.}
\begin{equation}
S=\frac{\Omega_{(D-2)} }{4G}\frac{r_{+}^{D-4}}{(1-\alpha)}L^2 \ .
\end{equation}
Moreover a straightforward computation shows that the thickness for this solution behaves like $r_{0}(\theta)=r_{+}\cos\theta$. However, since in this limit  $m\to0$, it follows from \eqref{E:iden1} that this implies $r_{+}\to0$ and thus $r_{0}\to0$. This is actually a prediction from the blackfold side since by taking Eq.~\eqref{E:ut} we see that in the case $\alpha=1$, $r_{0}=r_{+}$ except when $r\to r_{\rm max}$ in which case $r_{0}=0$, therefore such solution would only be regular if $r_{+}=0,\forall r$.

\subsubsection*{$0\le\alpha<1$: the ultra-spinning limit}

This limit resembles very closely that of the ultra-spinning MP black hole. To see this, start with the metric of the singly-spinning Kerr-AdS${}_{D}$ black hole  \cite{Hawking:1998kw} in spheroidal coordinates $(t,r,\theta,\phi,\Omega_{D-4})$ (see App.~\ref{ab}
for the multi-spin case). The ultra-spinning limit of the Kerr-AdS${}_D$
black holes is defined as
\begin{equation}
\label{newlim}
a\to\infty \spa L\to\infty \spa m\to\infty \ ,
\end{equation}
keeping $\alpha=a^2/L^2$ and $\hat{\mu}=2m/a^2$ fixed.
Consider in this limit the metric near the axis of rotation by defining a new coordinate $\sigma=a \sin\theta$ which remains finite as the axis is approached, i.e. as $\theta\to0$. Then the metric takes the form of
that of a flat black membrane
\begin{equation} \label{E:cylm1}
ds^2=-\left(1-\frac{\hat{\mu}}{r^{D-5}}\right)dt^2+\left(1-\frac{\hat{\mu}}{r^{D-5}}\right)^{-1}dr^2+r^2d\Omega_{D-4}^2+\frac{1}{1-\alpha}\left(d\sigma^{2}+\sigma^{2}d\phi^2\right) \ .
\end{equation}
The difference between this metric and the flat space case ($\alpha=0$) resides in the last term, which is multiplied by the factor $(1-\alpha)^{-1}$. In fact, since we are free to rescale the coordinate $\sigma$ by a factor of $\sqrt{(1-\alpha)^{-1}}$, we can eliminate the factor in front of the line-element of the two-plane $(\sigma,\phi)$. However we are only allowed to do this if $0\le\alpha<1$ since if $\alpha=1$ the metric diverges and if $\alpha>1$ the metric changes signature. In summary, the limit above is only valid if $\alpha$ lies within the range $0\le\alpha<1$ as  claimed in the discussion below \eqref{E:const2}.

 The physical properties of the Kerr-AdS${}_D$ solution in the limit
\eqref{newlim} can be easily obtained from \cite{Gibbons:2004ai}
\begin{equation} \label{E:um2}
M=\frac{\Omega_{(D-2)} }{8\pi G}\frac{\hat{\mu}}{(1-\alpha)^2}\left(1+\frac{(D-4)(1-\alpha)}{2}\right)a^2 \ ,
\end{equation}
\begin{equation}
S=\frac{\Omega_{(D-2)} }{4G}r_{+}^{D-4}\frac{a^2}{(1-\alpha)} \spa T=\frac{D-5}{4\pi r_+} \ ,
\end{equation}
\begin{equation} \label{E:uj2}
J=\frac{\Omega_{(D-2)} }{8\pi G}\frac{\hat{\mu}}{(1-\alpha)^2}a^3 \spa \Omega=\frac{1}{a} \ ,
\end{equation}
\begin{equation} \label{E:iden2}
r_{+}=\left(\frac{2m}{a^2}\right)^{\frac{1}{(D-5)}} \ .
\end{equation}
It is then seen that with the identification $\hat{r}_{+}=r_{+}$ and using $\Omega=a^{-1}$, we can reproduce from the blackfold approach (Eqs.~\eqref{E:ulm}-\eqref{E:ten}) the thermodynamic quantities given in \eqref{E:um2}-\eqref{E:uj2}. Furthermore, the thickness of this black membrane is given by $r_{0}(\theta)=r_{+}\cos\theta$. By looking at Eq.~\eqref{E:ut} and defining a new coordinate $\theta=\arcsin\left(r\Omega\sqrt{1-\alpha}\right)$ the two expressions for the thickness exactly match.

We note that the resulting metric does not represent an asymptotically flat solution. This is clear from the fact that there is a non-vanishing tension,
as seen in \eqref{E:ten}. Another way is by looking at the Quantum Statistical Relation that these black holes must satisfy,  this relation reads \cite{Gibbons:2004ai}
\begin{equation}
M-TS-\sum_{i}\Omega_{i}J_{i}=TI_{D} \ ,
\end{equation}
where the Euclidean action $I_{D}$ in this limit reduces to
\begin{equation}
I_{D}=\frac{1}{4T}\frac{\Omega_{(D-2)}}{(1-\alpha)}m \ .
\end{equation}
We can see that there is a factor of $(1-\alpha)^{-1}$ in the expression above, and one may check that only for $\alpha=0$ does one recover the Euclidean action for the asymptotically flat case.%
\footnote{To compare this result with the one obtained in \cite{Emparan:2009cs} note that the parameter $\mu$ in \cite{Emparan:2009cs} is related to $m$ by $\mu=2m$.}
 This limit thus represents an asymptotically AdS${}_{D}$ solution.

The existence of the ultra-spinning limit of  Kerr-AdS$_D$ black holes
described above provides non-trivial information on the stability properties
of these black holes. In the asymptotically flat case Ref.~\cite{Emparan:2003sy}
showed that ultra-spinning MP black holes  become membrane-like suggesting
that these should exhibit a GL-type instability \cite{Gregory:1993vy}%
\footnote{See \cite{Harmark:2007md} for a review on the GL instability.},
as confirmed in \cite{Dias:2009iu}.
Similarly, our analysis thus predicts that Kerr-(A)dS$_D$ black holes
for $D\geq 6$  suffer from an ultra-spinning GL-type instability when $\Omega L>1$.  This is in agreement with  the recent numerical analysis of Ref.~\cite{Dias:2010gk} for the singly-spinning case in AdS.
More generally, it follows From App.~\ref{ab} that in the multi-spin case
there is an ultra-spinning GL instability when $\Omega_i L>1$.

\subsection{Even-ball blackfolds in dS${}_{D}$ }

In this section we want to generalize the results of Sec.~\ref{S:eads} to a dS${}_{D}$ background. By performing a Wick rotation $L\to iL$ the action \eqref{E:acta} takes the same form but now with different functions $F(\rho),H(\rho)$ which transform accordingly. $Z=0$ is still a valid blackfold solution and the velocity field attains the velocity of light at a maximum value of
\begin{equation}
\rho_{\rm max}=2L(-L\Omega + \sqrt{L^2\Omega^2+1}) \ .
\end{equation}
There is thus no upper bound on the parameter $\alpha = (L^2 \Omega^2)^{-1}$ and hence $\alpha$ is free to take any value in the interval
\begin{equation}
\alpha\ge0 \ .
\end{equation}
In terms of $\alpha$ we can now distinguish two different regimes: \\
\textbf{\emph{(i)}} $\alpha=0$. This is the flat space case as noted previously in Sec.~\ref{S:eads}.\\
\textbf{\emph{(ii)}} $\alpha>0$. In this case $\rho_{max}\le2L$, and so the disc is cut at some value of $\rho$ in general but reaching the cosmological horizon when $\alpha\to\infty$ and hence $\Omega=0$ for which case the solution is static\footnote{We are grateful to Roberto Emparan for pointing this out to us.}.
As we will see below, this case ($\alpha$ >0) corresponds to the ultra-spinning limit of the Kerr-dS${}_{D}$ black hole.

We have not mentioned here any "ultra-spinning" regime. This is because the Kerr-dS${}_{D}$ does not show such special behavior when $a\to L$. To see this it suffices to look at Eq.~\eqref{E:div} and keep in mind that the ratio $\frac{l_{\parallel}}{l_{\perp}}\sim\Xi^{-\frac{D-1}{2(D-5)}}$ remains finite since now $\Xi=1+\alpha$, so that the event horizon does not pancake out along the plane of rotation.

The ultra-spinning limit of the Kerr-dS${}_{D}$ black hole can be obtained by performing the same Wick rotation on the metric \eqref{E:cylm1}
\begin{equation} \label{E:cylm2}
ds^2=-\left(1-\frac{\hat{\mu}}{r^{D-5}}\right)dt^2+\left(1-\frac{\hat{\mu}}{r^{D-5}}\right)^{-1}dr^2+r^2d\Omega_{D-4}^2+\frac{1}{1+\alpha}\left(d\sigma^{2}+\sigma^{2}d\phi^2\right) \ .
\end{equation}
It is then obvious that this metric is valid for all values of $\alpha\ge0$. Moreover, it is also a straightforward exercise to show that the physical properties of this solution matches those of the even-ball blackfold.
Finally, as in the AdS case, it follows that Kerr-dS${}_D$ black holes
for $D \geq 6$  have an ultra-spinning GL instability.

\section{Rings and helices} \label{S:st}

In \cite{Emparan:2009vd}  blackfold solutions were found in $D \geq 5$ with exotic horizons and a single axial $U(1)$ isometry.
These helical black rings and helical black strings constitute the first examples of asymptotically flat black holes that saturate%
\footnote{Ref.~\cite{Dias:2010eu} found evidence for another example,
in the context of time-independent perturbations at the onset of
instabilities of higher-dimensional black holes.}
the rigidity theorem \cite{Hollands:2006rj}. In this section we address the question whether helical rings can be attained as well in (A)dS${}_{D}$ spacetime. We will show that solutions describing helical rings with these symmetries, which are valid in the regime $r_{0}\ll L$, can also be constructed in these backgrounds. On the other hand the question wether or not helical strings can be constructed in these backgrounds remains open as it would require a different starting point from that of an asymptotically flat black brane (see Eq. \eqref{E:dsb}).

\subsubsection*{Helical black rings in (A)dS${}_{D}$}

In order to construct the action for these blackfolds it is convenient to write the metric of a $(D-1)$-dimensional spatially conformally flat subspace of (A)dS${}_{D}$ spacetime in such a way that all its $U(1)^{N}$ symmetries are explicit
\begin{equation} \label{E:adsS}
ds^2_{D-1} = H(\rho)^{-1}\left( \sum_{i=1}^{N}(\rho_{i}^{2} + \rho_{i}^{2}d\phi_{i}^{2})+\sum_{j=1}^{D-(2N+1)}dx_{j}^2\right) \spa \rho^{2}=\sum_{i=1}^{N}\rho_{i}^2+\sum_{j=1}^{D-(2N+1)}x_{j}^2 \ ,
\end{equation}
where we have the constraint $n\ge(2N-1)$. To embed the black 1-fold worldvolume $\mathcal{B}_{1}$ we set $x_{j}=0,\forall j$ and choose the set of scalars $\rho_{i}=\bar{R_{i}}$ and the spatial worldvolume coordinate $\sigma$ such that
\begin{equation} \label{E:embh}
\{\phi_{i}=n_{i}\sigma \spa 0\le\sigma\le2\pi,~i=1,...,N\} \ ,
\end{equation}
where we assume without loss of generality that $n_{i}\ge0$. The numbers $n_{i}$ must be integers in order for the ring to close up on itself and the smallest of them ($n_{\rm min}$) must be coprime with all the remaining ones to avoid multiple covering of the ring. The Killing vector field must be such that all the $U(1)^{N}$ symmetries generate the isometry of the worldvolume, giving
\begin{equation}
\textbf{k}=\frac{\partial}{\partial t}+\sum_{i=1}^{N}\Omega_{i}\frac{\partial}{\partial\phi_{i}} \ .
\end{equation}
The ratios between the angular velocities must be rational such that
\begin{equation}
\left|\frac{\Omega_{i}}{\Omega_{j}}\right|=\frac{n_{i}}{n_{j}},~\forall i,j \ ,
\end{equation}
and hence we can simply set $|\Omega_{i}|=\Omega n_{i}$. The action then takes the simple form
\begin{equation} \label{E:acth}
I[\{R\}]=2\pi R_{0}R\left(R_{0}^2-R^2\Omega^2\right)^{\frac{n}{2}} \ ,
\end{equation}
with $R_{0}^2=F(\rho)$, $R^2=H(\rho)^{-1}\sum_{i=1}^{N}n_{i}^2\bar{R_{i}}^2$ and $\rho^2=\sum_{i=1}^{N}\bar{R_{i}}^2$. The general
form of the action \eqref{E:acth} was given in Ref.~\cite{Emparan:2009cs}.
The action depends on the single scalar $R$ with $R_0$ a function of $R$, which should be taken into account when varying as well
as the fact that the variation should be orthogonal to the helix. As a result one single equation is found.

A more convenient action to work with can be obtained by making the following redefinition
\begin{equation}
R_{i}=\frac{\bar{R}_{i}}{1\pm\frac{\sum_{i=1}^{N}\bar{R_{i}}^2}{4L^2}} \ .
\end{equation}
The action \eqref{E:acth} then takes the same form but now with $R_{0}^2=\mathcal{V}(r)$, $R^2=\sum_{i=1}^{N}n_{i}^2R_{i}^2$ and $r^2=\sum_{i=1}^{N}R_{i}^2$.
Varying this with respect to $R$ leads to \cite{Emparan:2009vd}
\begin{equation}\label{E:1f}
\Omega^2=\frac{R_{0}^2}{R^2} \frac{1+(n+1)\frac{dlnR_{0}}{dlnR}}{n+1+\frac{dlnR_{0}}{dlnR}} \ .
\end{equation}
In the case at hand for AdS${}_D$ the solution becomes
\begin{equation}
\label{omegahel}
\Omega^{2}=\frac{1+\textbf{R}^2}{R^2}\frac{(1+\textbf{R}^2)(n+2)-(n+1)}{(1+\textbf{R}^2)(n+2)-1} \ ,
\end{equation}
where we have defined $\textbf{R}^2=L^{-2}\sum_{i=1}^{N}R_{i}^2$. This agrees with the result for planar black rings $n_{i}=1,~\forall i$ of \eqref{E:oms} with $p=1$. The equilibrium condition for helical rings and planar rings is exactly the same but with a more complicated expression for $R$ in the former case. The only difference resides in the fact that in the planar case specifying $R$ immediately specifies $R_{0}$ for these backgrounds while for the helical case one needs to specify $R$ and $R_{0}$ independently since the latter is a function of $\sum_{i=1}^{N}R_{i}^2$. Note that it follows from \eqref{omegahel}
 that static helical black rings can exist in dS${}_{D}$ provided
\begin{equation}
\label{rstat}
\textbf{R}^2=\frac{1}{n+2} \ ,
\end{equation}
which is the same condition as for static planar rings in dS${}_{D}$ and hence independent of the integers $n_{i}$. Accordingly, \eqref{rstat} can also be obtained from \eqref{E:conds} in the special case of $p=1$.

We now proceed  by describing the physical quantities of the helical AdS black
rings
\begin{equation}
M=\frac{\Omega_{(n+1)}}{8 G}(n+2)r_{0}^{n}(1+\textbf{R}^2)^{\frac{3}{2}}\sqrt{\sum_{i=1}^{N}n_{i}^2R_{i}^2} \ ,
\end{equation}
\begin{equation}
J_{i}=\pm\frac{\Omega_{(n+1)}}{8 G}r_{0}^{n}((1+\textbf{R}^2)(n+2)-1)\sqrt{1-\frac{n}{(1+\textbf{R}^2)(n+2)-1}}n_{i}R_{i}^2 \ ,
\end{equation}
\begin{equation}
S=\frac{\pi\Omega_{(n+1)}}{2 G}r_{0}^{n+1}\sqrt{\frac{(1+\textbf{R}^2)(n+2)-1}{n}}\sqrt{\sum_{i=1}^{N}n_{i}^2R_{i}^2} \ .
\end{equation}
These quantities agree with the ones computed in \cite{Emparan:2009vd} for helical rings in flat space (when taking $L \rightarrow \infty$) and with the ones computed in \cite{Caldarelli:2008pz} for planar rings in (A)dS (when taking $n_i=1,~ \forall i$). For completeness we also give the tension for these helical rings
\begin{equation}
\mathcal{T}=-\frac{\Omega_{(n+1)}}{8 G}(n+2)r_{0}^{n}(1+\textbf{R}^2)^{\frac{1}{2}}\textbf{R}^2\sqrt{\sum_{i=1}^{N}n_{i}^2R_{i}^2} \ .
\end{equation}
As expected, the tension vanishes only in the asymptotically flat case when $\textbf{R}\to0$. It can be shown that these physical properties satisfy the Smarr relation \eqref{smarr} and the relation \eqref{Grel}.

\subsubsection*{Helical rings in different backgrounds}

We would now like to give a few comments on helical rings in different background geometries. In fact it seems likely that helical rings can exist in any spherically symmetric background of the form \eqref{E:dsads1} since these can always be put into coordinates for which the potential $\mathcal{V}(r)$ is constant along the ring. As a matter of a fact, Eq.~\eqref{E:1f} first derived in \cite{Emparan:2009vd} holds for any $1$-fold assuming only the existence of a background timelike and spacelike Killing vector, hence a valid solution should exist for such backgrounds. This leads us to the following conjecture
\begin{conjecture}
Neutral helical black ring solutions exist in any background with spherical symmetry in the regime $r_{0}\ll |\Lambda|^{-\frac{1}{2}}$.
\end{conjecture}
As an example of a different spherically symmetric background we take the Schwarzschild-Tangherlini solution in $D$ dimensions (Sch${}_{D}$) as the background and try to construct a helical black Saturn.%
\footnote{The black Saturn solution in five dimensions was constructed in
Ref.~\cite{Elvang:2007rd}. See also Refs.~\cite{Emparan:2007wm,Caldarelli:2008pz,Emparan:2009vd,Emparan:2010sx}
for results on black Saturns in higher dimensions.}
The Sch${}_{D}$ metric can be written as in \eqref{E:dsads1} but with
\begin{equation}\label{E:pots}
\mathcal{V}(r)=1-\left(\frac{\mu}{r}\right)^{D-3} \spa \mu\le\ r\le\infty \ .
\end{equation}
By performing the transformation  $r=(1+\frac{\mu^{D-3}}{4\rho^{D-3}})^{\frac{2}{D-3}}\rho$ one can bring the Schwarzschild metric to the form \eqref{E:dsads2} with
\begin{equation}
F(\rho)=\frac{(1-\frac{\mu^{D-3}}{4\rho^{D-3}})^2}{(1+\frac{\mu^{D-3}}{4\rho^{D-3}})^2}
\spa H(\rho)=(1+\frac{\mu^{D-3}}{4\rho^{D-3}})^2
\spa \left(\frac{5}{4}\mu^{\frac{D-3}{2}}\right)^{\frac{2}{D-3}}\le \rho\le\infty \ .
\end{equation}
Using the embedding \eqref{E:embh} the action reduces to \eqref{E:acth} but with $R_{0}^2=\mathcal{V}(r)$ given by \eqref{E:pots}. The solution can be obtained from \eqref{E:1f} and reads
\begin{equation}
\Omega^{2}=\frac{(1-\mathbf{m}^{n+1})}{R^2}\frac{(1-\mathbf{m}^{n+1})(2-(n+1)^2)+(n+1)^2}{(n+1)\mathbf{m}^{n+1}} \ ,
\end{equation}
where we have defined the parameter $\textbf{m}=\frac{\mu}{\sum_{i}^{N}R_{i}^2}$ and used the fact that in this case $D=n+4$. One can go even further and perform the same calculation for a general potential of the form $\mathcal{V}(r)$ with $r^2=\sum_{i=1}^{N}R_{i}^2$, the equilibrium condition for $\Omega$ is given by the relation
\begin{equation}
\Omega^2=\frac{R_{0}^2}{R^2}\frac{2R_{0}^2+(n+1)R_{0}^{2'}\sqrt{\sum_{i=1}^{N}R_{i}^2}}{2R_{0}^2(n+1)+R_{0}^{2'}\sqrt{\sum_{i=1}^{N}R_{i}^2}} \ ,
\end{equation}
where $R_{0}^{2}=\mathcal{V}(r)$ and $R_{0}^{2'}\equiv \partial_{r}R_{0}^2$. This generalizes the equilibrium condition obtained in \cite{Caldarelli:2008pz} for planar rings in backgrounds of this form.

\section{Concluding remarks and future directions}

Throughout the course of this work we have used the blackfold approach to scan for possible neutral black hole solutions in (A)dS${}_{D}$ spacetime with new horizon topologies. We emphasize that the blackfold equations \eqref{E:ext}-\eqref{E:bfe} that we have solved are the zeroth order equations in which the
black brane is treated at the probe level. Using the method of matched asymptotic
expansion (MAE) one may compute higher-order corrections in a perturbative series.
It may happen that backreaction makes it impossible for a leading-order solution
to remain stationary, and, with our present knowledge, this has to be examined
on a case-by-case basis.

 The next-to-leading order analysis  has been successfully performed for black rings \cite{Emparan:2007wm} and black tori \cite{Emparan:2009vd} in asymptotically flat space and black rings in (A)dS space \cite{Caldarelli:2008pz}. In fact, following \cite{Emparan:2009vd} the analysis of \cite{Caldarelli:2008pz} is probably easily adapted to compute the
next-to-leading order correction of the black tori in (A)dS obtained in
this paper (i.e. the case $p_a=1, \, \forall a$ in  Sec.~\ref{sec:prod}).
It is an interesting open problem to do this for
the higher odd-sphere solutions of \cite{Emparan:2009vd} and those in (A)dS obtained in Sec.~\ref{S:odd}.
Finally, we note that further evidence for the regularity of the solutions presented here follows from the fact that we have shown that the even-ball blackfolds reproduce exactly the thermodynamical properties of certain limits of the Kerr-(A)dS${}_{D}$ black hole, which is in itself a regular solution.

On the other hand, the usefulness of the blackfold approach also resides in
its ability to exclude possible horizon topologies, at least in the regime
of widely separated scales in which the effective theory is valid.
Thus, when a particular embedding does not solve the blackfold equations one
 may straight away affirm that such horizon geometry cannot be that of a solution of Einstein equations, in the theory and regime under consideration, without any need of checking further corrections.  For this reason the blackfold method is an excellent tool for
probing new stationary black hole solutions, since besides giving a space of possible geometries, it allows one to exclude with little effort a large number of other topologies.%
\footnote{Possible horizon topologies have been considered from alternate
points of view in \cite{Galloway:2005mf} and \cite{Kleihaus:2009wh}.}

These `impossible' geometries, when determined, are truths of the theory and hence can be stated as theorems. An example of this is the uniform black cylinder found in \cite{Emparan:2009vd}, which, can be easily shown not to be a solution of the blackfold equations in global (A)dS. But not only can the blackfold approach ascertain such statements, it can also provide a great deal of intuition as to why certain geometries are not possible and as to which ones are more likely to exist. In the case of the uniform black cylinder, the physical reason for it not being a solution is that, contrary to Minkowski space, in global (A)dS there is no extra translational symmetry. However, the construction of non-uniform black cylinders is a rather more difficult task to accomplish in these backgrounds and remains an open problem.

We also would like to give a few remarks on the geometric character of this methodology. A blackfold solution provides a possible geometry for the event horizon of a black hole spacetime. Bearing this in mind, it is then not surprising that there could be different spacetimes which in some appropriate limit could be captured by the same blackfold solution. However, a blackfold solution is not merely of geometric character as it is also supplemented with well defined physical properties, hence a given spacetime has to not only match its geometric properties but also its physical ones. An example of this is the construction of even-ball (A)dS blackfolds in Sec.~\ref{S:even} where to leading order the horizon is extended along the plane $z=0$ and we could identify it according to the values of the parameter $\alpha$, with an ultra-spinning and an "ultra-spinning" limit of the Kerr-(A)dS${}_{D}$ black hole.

We note that it does not follow from the above that these are necessarily the only two possible identifications that one can make with these even-ball blackfolds. As a matter of fact, the case $\alpha=1$ could in principle have another possible identification, namely that of another limit of the Kerr-(A)dS${}_{D}$ spacetime
which near the axis of rotation looks like a rotating black hyperboloid membrane \cite{Caldarelli:2008pz}, with metric
\begin{equation} \label{E:ultrah}
ds^2=-f(r)(dt-L\sinh^2(\sigma/2)d\phi)^2+\frac{dr^2}{f(r)}+\frac{L^2}{4}
\left(1+\frac{r^2}{L^2}\right)(d\sigma^2+\sinh^2\sigma d\phi^2)+r^2d\Omega_{D-4}^2 \ ,
\end{equation}
where
\begin{equation}
f(r)=1-\frac{2m}{r^{D-5}(r^2+L^2)}+\frac{r^2}{L^2} \ .
\end{equation}
This is the metric obtained by taking $a\to L$ as in the "ultra-spinning" case but instead of requiring the mass to be finite, one requires the horizon size to be finite. This means that the physical properties of this solution are the same as in the "ultra-spinning" case \eqref{E:um1}-\eqref{E:iden1} but now the parameter $\hat{\mu}$ diverges. The BPS bound $J\le ML$ is not violated since both $J$ and $M$ diverge with a constant ratio $J/ML\to1$. For  small values of $m$ (and hence  $r_{+}\ll L$ with $r_{+}=\frac{2m}{L^2}$)  the metric above (for $D\ge6$) actually reduces to the metric \eqref{E:ultra1} near $\sigma=0$. However, the thickness for this solution varies as $r_{0}(\sigma)=r_{+}\cos\theta$ with $r_{+}\ne0$ since now $m\ne0$.
But we have seen from \eqref{E:ut} that the thickness must remain constant and equal to zero if $\alpha=1$, so that this limit cannot be captured by the same blackfold solution.

This could have been foreseen since the metric \eqref{E:ultrah} is a solution of Einstein equations with a cosmological constant, meaning that at least near the axis the Kerr-(A)dS${}_D$ solution in this limit is not locally flat, so it can not be reproduced by the blackfold approach using \eqref{E:dsb} as the starting point. Nevertheless, applying  MAE to the case under consideration would then give a metric which according to a set of parameters would describe the two distinct ultraspinning limits, and since all these regimes found here are limits of the Kerr-(A)dS${}_{D}$ black hole the resulting metric from the MAE procedure would most likely describe \textbf{\emph{(i)}} a specific sector of the Kerr-(A)dS${}_{D}$ spacetime or \textbf{\emph{(ii)}} a more general solution in the (A)dS${}_{D}$ background which englobes all the possible limits that reproduce the geometric and physical properties of these blackfolds. Such investigation would be a worthy endeavor.

We finally mention briefly several other open problems that deserve
further study. First of all, following \cite{Emparan:2007wm,Caldarelli:2008pz,Emparan:2010sx}
it would be interesting to study in more detail the connection between
different phases for black objects in (A)dS. Furthermore the possibility
of describing large black objects in (A)dS with the blackfold approach would
be important to pursue. Another open direction would be to consider charged
blackfolds in (A)dS using the methods developed in \cite{upcoming} and
\cite{upcoming2}.
We also stress that the utility of the blackfold approach is not only restricted to describing novel classes of stationary black holes, but also enables analyzing their dynamics (stability and time-evolution) in specific physical situations \cite{Emparan:2009at,Camps:2010br}. This would be worthwhile to examine
further for (A)dS blackfolds. In particular, it would be interesting to
analyze in more detail using the blackfold approach the GL instability predicted from our results for the ultra-spinning Kerr-(A)dS${}_{D}$ black holes.

\section*{Acknowledgments}

Jay would like to thank FCT Portugal for the grant SFRH/BD/45893/2008. Jay would also like to thank the Auroville community, India, where part of his work was done, for all the conversations on consciousness and the meaning of life and to the Christiania community, Copenhagen, where part of this work was put into words, for its lively activist spirit. We would also like to thank Troels Harmark, Vasilis Niarchos, Konstantinos Zoubos and especially Roberto Emparan  for useful discussions.

\appendix

\section{Multi-spin Kerr-(A)dS${}_{D}$ black holes as blackfolds} \label{as}

In this appendix we generalize the even-ball (A)dS blackfold construction of Sec.~\ref{S:eads}, which focused on the case of a spinning disc, to include the multi-spinning case.

In this case the $(2k)$-ball rotates rigidly in $k$ independent
two-planes and the total velocity is given by
\begin{equation}
V=\left(1+\frac{\sum_{i=1}^{N}\rho_{i}^2}{4L^2}\right)^{-1}\sqrt{\sum_{i=1}^{N}\rho_{i}^2\Omega_{i}^2} \ .
\end{equation}
The boundary of the ball is given by the locus where $V=1$,
which is solved by
\begin{equation}
\sum_{i=1}^{k}\Omega_{i}^2r_{i}^2(1-\alpha_{i})=1 \ ,
\end{equation}
where we have defined $r_{i}=\rho_{i}/(1-\frac{\sum_{i}^{N}\rho_{i}^2}{4L^2})$ and $\alpha_{i}=(\Omega_{i}^2L^2)^{-1}$. According to the parameters $\alpha_{i}$ we can distinguish three different cases: \\
\textbf{\emph{(i)}} $\alpha_{i}=0,\forall i$. This corresponds to the
even-ball blackfold construction of the ultra-spinning MP black hole constructed and discussed in Refs.~\cite{Emparan:2009cs,Emparan:2009vd}.\\
\textbf{\emph{(ii)}} $\alpha_{i}=1,\forall i$. This corresponds to the "ultra-spinning" limit $\Omega_{i}\to L^{-1}$ given in \cite{Caldarelli:2008pz}.  \\
\textbf{\emph{(iii)}} $0<\alpha_{i}<1,\forall i$. This corresponds to a new ultra-spinning limit of the Kerr-AdS${}_{D}$ black hole presented
in App.~\ref{ab}.

The physical properties of the even-ball blackfolds can be computed from \eqref{E:m}-\eqref{E:t} and read
\begin{equation} \label{E:aum1}
M=\frac{\Omega_{(D-2)}\hat{r}_{+}^n}{8\pi G \prod_{j}(1-\alpha_{j})}\left(\sum_{j=1}^{k}\frac{1}{1-\alpha_{j}}+\frac{n+1}{2}\right)\prod_{j}\frac{1}{\Omega_{j}^2} \ ,
\end{equation}
\begin{equation}
J_{i}=\frac{\Omega_{(D-2)}\hat{r}_{+}^n\prod_{j}\frac{1}{\Omega_{j}^2}}{8\pi G\prod_{j}(1-\alpha_{j})}\frac{1}{1-\alpha_{i}}\frac{1}{\Omega_{i}} \ ,
\spa
\label{E:aut1}
S=\frac{\Omega_{(D-2)}\hat{r}_{+}^{n+1}}{4G}\prod_{j}\left(\frac{1}{1-\alpha_{j}}\frac{1}{\Omega_{j}^2}\right) \end{equation}
\begin{equation}
\mathcal{T}=-\frac{\Omega_{(D-2)}\hat{r}_{+}^n}{8\pi G \prod_{j}(1-\alpha_{j})}\left(\sum_{j=1}^{k}\frac{1}{1-\alpha_{j}}-k\right)\prod_{j}\frac{1}{\Omega_{j}^2} \ ,
\end{equation}
where $\hat{r}_+ = n/2\kappa$.
These expressions reduce to the results \eqref{E:ulm}-\eqref{E:ten}
for the for the singly-spinning case.
The quantities above can be shown to satisfy the Smarr relation \eqref{smarr}
and \eqref{Grel}.
The horizon thickness $r_{0}$ is given by
\begin{equation}
r_{0}=\frac{n}{2\kappa}\sqrt{1-\sum_{i=1}^{k}\Omega_{i}^2r_{i}^2(1-\alpha_{i})} \ .
\end{equation}
We can identify these even-ball blackfolds with two different limits of the  Kerr-AdS${}_{D}$  as follows:

\subsubsection*{$\alpha_{i}=1,~\forall i$: the "ultra-spinning" limit}

A detailed study of this limit has been presented in Ref.~\cite{Caldarelli:2008pz}. The limit amounts to taking $a_{i}\to L,~\forall i$ while keeping $\hat{\mu}\equiv\frac{2m}{L^{2k}\prod_{i=1}^{k}\Xi_{i}}$ finite. The resulting metric gives a flat membrane with metric as in \eqref{E:ultra1} where the horizon size $r_{+}$ shrinks to zero. Its physical properties are the same as for the ultra-spinning case \eqref{E:am1}-\eqref{E:as1} in the limit $r_{+}\ll L$ but now with $a_{i}=L,~\forall i$. With the identifications
\begin{equation}
\Omega_{i}=\frac{1}{L},~\forall i \spa
\hat{r}_{+}=r_{+}=\left(\frac{2m}{L^{2k}}\right)^{\frac{1}{D-2k-3}} \ ,
\end{equation}
the physical properties of this solution exactly match those of the blackfolds \eqref{E:aum1}, \eqref{E:aut1}.

\subsubsection*{$0\le\alpha_{i}<1,~\forall i$: the ultra-spinning limit}

A careful calculation of this limit is given in App.~\ref{ab}. It amounts to taking $k$ number of spins to infinity while keeping the ratios $\alpha_{i}=\frac{a_{i}^2}{L^2}$ and $\hat{\mu}=\frac{2m}{\prod_{i=1}^{k}a_{i}^2}$ finite. The resulting metric is presented in \eqref{E:aultra2} and has the geometry of a flat black membrane near the axes of rotation. With the identifications
\begin{equation}
\Omega_{i}=\frac{1}{a_{i}},~\forall i \spa
\hat{r}_{+}=r_{+}=\left(\frac{2m}{\prod_{i=1}^{k}a_{i}^2}\right)^{\frac{1}{D-2k-3}} \ ,
\end{equation}
the physical properties \eqref{E:am1}-\eqref{E:as1} match precisely those of the blackfolds \eqref{E:aum1}, \eqref{E:aut1}.

\section{The ultra-spinning Kerr-(A)dS${}_{D}$ black hole} \label{ab}

In this appendix we  take the ultra-spinning limit of the Kerr-(A)dS${}_{D}$ black hole with an arbitrary number of spins in $D\ge6$. We focus here on the AdS case since we can obtain its dS counterpart by performing a Wick rotation.

Defining $N=\frac{D-1}{2}\, {\rm mod}\, 2$ as the number of two-planes, $\epsilon$ by the relation $D=2N+1+\epsilon$, with $\epsilon$ being either $1$ for even dimensions or $0$ for odd dimensions, and introducing $N+\epsilon$ direction cosines $\mu_{i}$ obeying the constraint $\sum_{i=1}^{N+\epsilon}\mu_{i}^2=1$, the metric of this spacetime can be conveniently written in Boyer-Lindquist coordinates as \cite{Gibbons:2004uw}
\begin{equation}
\begin{split}
ds^2=& -W(1+\frac{r^2}{L^2})dt^2+\frac{2m}{U}\left(Wdt-\sum_{i=1}^{N}\frac{a_{i}\mu_{i}^2d\varphi_{i}}{\Xi_{i}}\right)^2+\sum_{i=1}^{N}\frac{r^2+a_{i}^2}{\Xi_{i}}\mu_{i}^2d\varphi_{i}^2\\
&+\frac{Udr^2}{V-2m}+\sum_{i=1}^{N+\epsilon}\frac{r^2+a_{i}^2}{\Xi_{i}}d\mu_{i}^2-\frac{1}{L^2W(1+\frac{r^2}{L^2})}\left(\sum_{i=1}^{N+\epsilon}\frac{r^2+a_{i}^2}{\Xi_{i}}\mu_{i}d\mu_{i}\right)^2 \ ,
\end{split}
\end{equation}
where
\begin{equation}
W=\sum_{i=1}^{N+\epsilon}\frac{\mu_{i}^2}{\Xi_{i}}
\spa U=r^{\epsilon}\sum_{i=1}^{N+\epsilon}\frac{\mu_{i}^2}{r^2+a_{i}^2}\prod_{w=1}^{N}(r^2+a_{w}^2) \ ,
\end{equation}
\begin{equation}
V=r^{\epsilon-2}(1+\frac{r^2}{L^2})\prod_{i=1}^{N}(r^2+a_{i}^2)
\spa \Xi_{i}=1-\frac{a_{i}^2}{L^2} \ .
\end{equation}
The horizon $r_{+}$ sits at the largest positive real root of $V-2m=0$. We assume that the parameters are chosen in a such a way that a horizon exists. In fact a horizon always exists for odd $D$ if any two of the spin parameters $a_{i}$ vanish, while for even $D$ its existence is guaranteed if only one vanishes.

We now take an $k$ number of spin parameters $a_{i}$ to be very large as compared to the remaining $N-k$ ones, i.e.,
\begin{equation}
\begin{split}
&a_{j}\to\infty,~j=1,...,k \ , \\
&a_{l}~{\rm finite},~l=k+1,...,N \ .
\end{split}
\end{equation}
Furthermore we take $L$ to be of an equal magnitude as compared to the $a_{j}$ parameters such that the ratios $\alpha_{j}=a_{j}^2/L^2$ remain constant, i.e.
\begin{equation}
\begin{split}
&L\to\infty \ , \\
\Xi_{j}\to&1-\alpha_{j}\spa \Xi_{l}\to0 \ .
\end{split}
\end{equation}
Moreover we take $m\to\infty$ such that the ratio $\hat{\mu}=\frac{2m}{\prod_{j}a_{j}^2}$ remains finite and define new coordinates $\sigma_{j}=a_{j}\mu_{j}$ that remain finite as we approach $\mu_{j}\to0$. The remaining $\mu_{l}$ stay finite and satisfy $\sum_{l}\mu_{l}^2+\mu_{N+1}^2=1$. In this limit we find that the metric function $W$ behaves like $W\to1$ while the remaining metric functions become
\begin{equation}
\begin{split}
&U\to r^{\epsilon}\frac{\mu_{l}^2}{r^2+a_{l}^2}\prod_{l}(r^2+a_{l}^2)^2\prod_{j}a_{j}^2\equiv r^{\epsilon}\hat{F}\hat{\Pi}\prod_{j}a_{j}^2 \ ,\\
&V\to r^{\epsilon-2}\prod_{l}(r^2+a_{l}^2)^2\prod_{j}a_{j}^2\equiv r^{\epsilon-2}\hat{\Pi}\prod_{j}a_{j}^2 \ ,\\
\end{split}
\end{equation}
where we have assumed the summation convention over $j$ and $l$. The limiting metric then reads
\begin{equation} \label{E:aultra2}
\begin{split}
ds^2=&-dt^2+r^2d^2\mu_{N+1}+(r^2+a_{l}^2)(d\mu_{l}^2+\mu_{l}^2d\varphi_{l}^2)+\frac{\hat{\mu}r^{-\epsilon}}{\hat{F}\hat{\Pi}}(dt+a_{l}\mu_{l}^2\varphi_{l}^2)^2+\frac{r^2\hat{F}\hat{\Pi}}{\hat{\Pi}-\hat{\mu}r^{2-\epsilon}}dr^2\\
&+\frac{1}{1-\alpha_{j}}(d\sigma_{j}^2+\sigma_{j}^2\varphi_{j}^2) \ .
\end{split}
\end{equation}
This limit looks just like the ultra-spinning limit of the MP black hole \cite{Emparan:2003sy} except for the extra factor $(1-\alpha_{j})^{-1}$ in the
last term. Indeed since the parameters $\alpha_{j}$ must lie within the range $0\le\alpha_{j}<1$, since otherwise the metric either changes signature or diverges, we can rescale $\sigma_{j}$ such that $\sigma_{j}\to\hat{\sigma_{j}}= \sqrt{(1-\alpha_{j})^{-1}}\sigma_{j}$, and hence the metric above will be that of a rotating black $2k$-brane with rotation along the spherical $S^{D-(2k+1)}$ sections of the horizon.

The physical properties of this solution can be easily computed from \cite{Gibbons:2004ai} with the result
\begin{equation} \label{E:am1}
M=\frac{\Omega_{(D-2)}\hat{\mu}}{8\pi G \prod_{j}(1-\alpha_{j})}\left(\sum_{j=1}^{k}\frac{1}{1-\alpha_{j}}-\frac{D-2k-1}{2}\right)\prod_{j}a_{j}^2 \ ,
\end{equation}
\begin{equation}
S=\frac{\Omega_{(D-2)}r_{+}^{2(N-k)-1+\epsilon}}{4G}\prod_{j}\frac{a_{j}^2}{1-\alpha_{j}^2}
\spa T=\frac{2(N-k-1)+\epsilon}{4\pi r_{+}} \ ,
\end{equation}
\begin{equation}
\label{E:as1}
J_{i}=\frac{\Omega_{(D-2)}\hat{\mu}\prod_{j}a_{j}^2}{8\pi G\prod_{j}(1-\alpha_{j})}\frac{a_{i}}{1-\alpha_{i}} \spa
\Omega_{i}=\frac{1}{a_{i}} \ ,
\end{equation}
\begin{equation}
\label{E:as2}
 r_{+}=\left(\frac{2m}{\prod_{j}a_{j}^2}\right)^{\frac{1}{D-2k-3}}
\spa i=1,...,k \ .
\end{equation}
These correctly reduce to the results \eqref{E:um2}-\eqref{E:iden2} for the singly-spinning case.
The quantities in \eqref{E:am1}-\eqref{E:as2} obey the Quantum Statistical Relation
\begin{equation}
M-\sum_{j}\Omega_{j}J_{j}-TS=\frac{\Omega_{(D-2)}}{8\prod_{j}(1-\alpha_{j})}\hat{\mu}\prod_{j}a_{j}^2 \ ,
\end{equation}
and thus only when $\alpha_{j}=0,\forall j$ does one recover the flat space result.
%Hence this limit still describes an AdS${}_{D}$ solution and is not asymptotically flat even though we have taken $L\to\infty$.

We conclude this appendix by noting that there are only three ways of making the Kerr-AdS${}_{D}$ black hole ultra-spin in the sense that $J\to\infty$. This is obvious by looking at the general expression for $J_{i}$
\begin{equation}
J_{i}=\frac{\Omega_{(D-2)}m}{4\pi\prod_{w}^{N}\Xi_{w}}\frac{a_{i}}{\Xi_{i}} \ .
\end{equation}
One way is to send $a_{i}\to\infty$ but keeping $\Xi_{i}$ finite since otherwise $J_{i}\to0$. This can only be done by simultaneously sending $L\to\infty$. Another way is to send $a_{i}\to L$ leading to $\Xi_{i}\to0$, if then one sends $m\to0$, $J_{i}$ remains finite and we obtain the "ultra-spinning" limit of \eqref{E:ultra1}. If otherwise we keep $m$ finite we obtain the rotating hyperboloid membrane of \eqref{E:ultrah}. This limit has the consequence that the horizon size is kept finite but the mass diverges, i.e., $M\to\infty$ and hence it is not ultra-spinning in the sense that $J \gg M$. A third possibility would be to naively send $m\to\infty$ but this results in a black hole with infinite horizon radius, which is senseless unless we simultaneously send $a_{i}\to\infty$ and $L\to\infty$ and thus recover the limit taken here. Moreover, since we have taken $L\to\infty$ while keeping $r_{+}$ finite, the ultra-spinning limit of the Kerr-(A)dS black hole exists only in the regime $r_{+}\ll L$ and hence can be fully captured by the blackfold approach.

\addcontentsline{toc}{section}{References}

\providecommand{\href}[2]{#2}\begingroup\raggedright\endgroup

%\bibliographystyle{newutphys}
%\bibliography{bibrings}
\end{document}